\documentclass[]{aastex631}

\shorttitle{\ce{CH3OH} formation from $\ce{H2CO} + \ce{CH3O}$: a dominant route under dense-molecular-cloud conditions}
\shortauthors{Santos et al.}
\graphicspath{{./}{figures/}}

\usepackage{graphicx}
\usepackage[version=4]{mhchem}
\usepackage{units}
\usepackage{amsmath,amssymb}
\usepackage[utf8]{inputenc}
\usepackage[T1]{fontenc}
\usepackage{xcolor}
\usepackage{appendix}
\usepackage{booktabs}
\usepackage{multirow}

\usepackage{todonotes}

\begin{document}

\title{First experimental confirmation of the \ce{CH3O} + \ce{H2CO} $\to$ \ce{CH3OH} + \ce{HCO} reaction: expanding the \ce{CH3OH} formation mechanism in interstellar ices}

\author[0000-0002-3401-5660]{Julia C. Santos}
\affiliation{Laboratory for Astrophysics, Leiden Observatory, Leiden University, PO Box 9513, 2300 RA Leiden, The Netherlands}

\author[0000-0001-6877-5046]{Ko-Ju Chuang}
\affiliation{Laboratory for Astrophysics, Leiden Observatory, Leiden University, PO Box 9513, 2300 RA Leiden, The Netherlands}

\author[0000-0001-6705-2022]{Thanja Lamberts}
\affiliation{Leiden Institute of Chemistry, Gorlaeus Laboratories, Leiden University, PO Box 9502, 2300 RA Leiden, The Netherlands}
\affiliation{Laboratory for Astrophysics, Leiden Observatory, Leiden University, PO Box 9513, 2300 RA Leiden, The Netherlands}

\author[0000-0003-2434-2219]{Gleb Fedoseev}
\affiliation{Research Laboratory for Astrochemistry, Ural Federal University, Kuibysheva St. 48, 620026 Ekaterinburg, Russia}

\author[0000-0002-2271-1781]{Sergio Ioppolo}
\affiliation{School of Electronic Engineering and Computer Science, Queen Mary University of London, Mile
End Road, London E1 4NS, UK}

\author[0000-0002-8322-3538]{Harold Linnartz}
\affiliation{Laboratory for Astrophysics, Leiden Observatory, Leiden University, PO Box 9513, 2300 RA Leiden, The Netherlands}




\begin{abstract}

The successive addition of \ce{H} atoms to \ce{CO} in the solid phase has been hitherto regarded as the primary route to form methanol in dark molecular clouds. However, recent Monte Carlo simulations of interstellar ices alternatively suggested the radical-molecule H-atom abstraction reaction $\ce{CH3O} + \ce{H2CO} \to \ce{CH3OH} + \ce{HCO}$, in addition to $\ce{CH3O} + \ce{H} \to \ce{CH3OH}$, as a very promising and possibly dominating (70--90\%) final step to form \ce{CH3OH} in those environments. Here, we compare the contributions of these two steps leading to methanol by experimentally investigating hydrogenation reactions on \ce{H2CO} and \ce{D2CO} ices, which ensures comparable starting points between the two scenarios. The experiments are performed under ultrahigh vacuum conditions and astronomically relevant temperatures, with H:\ce{H2CO} (or \ce{D2CO}) flux ratios of 10:1 and 30:1. The radical-molecule route in the partially deuterated scenario, $\ce{CHD2O} + \ce{D2CO} \to \ce{CHD2OD} + \ce{DCO}$, is significantly hampered by the isotope effect in the D-abstraction process, and can thus be used as an artifice to probe the efficiency of this step. We observe a significantly smaller yield of $\ce{D2CO} + \ce{H}$ products in comparison to $\ce{H2CO} + \ce{H}$, implying that the \ce{CH3O}-induced abstraction route must play an important role in the formation of methanol in interstellar ices. Reflection-Absorption InfraRed Spectroscopy (RAIRS) and Temperature Programmed Desorption-Quadrupole Mass Spectrometry (TPD-QMS) analyses are used to quantify the species in the ice. Both analytical techniques indicate constant contributions of $\sim$80\% for the abstraction route in the 10-16 K interval, which agrees well with the Monte Carlo conclusions. Additional \ce{H2CO} + \ce{D} experiments confirm these conclusions.

\end{abstract}



\section{Introduction} \label{sec:intro}

Methanol (\ce{CH3OH}) is abundantly detected in interstellar environments and is one of the main components of interstellar ices. Its formation is tightly connected to the evolution of molecular clouds. In these dense and cold regions, CO molecules present in the gas phase freeze-out and form an apolar coating on top of icy grains, largely comprising of \ce{H2O} and \ce{CO2} ice. Methanol is then efficiently formed through the hydrogenation of CO in the solid phase \citep{Tielens1982,Charnley1992,Hiraoka1994,Watanabe2002,Fuchs2009}. The presence of methanol in CO-rich ice is supported by a series of comparisons between laboratory and observational evidence (e.g., \citealt{Bottinelli2010,Cuppen2011, Penteado2015}). Furthermore, as opposed to CO hydrogenation, neither gas-phase nor alternative solid-phase routes can explain the observed \ce{CH3OH} abundances \citep{Geppert2005,Garrod2006,Watanabe2007}. As a secondary mechanism, it has also been shown to form in \ce{H2O}-rich ices prior to the heavy \ce{CO} freeze-out---thus in an earlier evolutionary stage of the cloud \citep{Bergner2017,Qasim2018,Potapov2021,Molpeceres2021}.

So far, the main proposed hydrogenation route to form methanol is through successive addition reactions of H atoms to CO ice, as follows:
\begin{equation}
    \ce{CO} + \ce{H} \to \ce{HCO}
    \label{eq:CO+H}
\end{equation}
\begin{equation}
    \ce{HCO} + \ce{H} \to \ce{H2CO}
    \label{eq:HCO+H}
\end{equation}
\begin{equation}
    \ce{H2CO} + \ce{H} \to \ce{CH3O}
    \label{eq:H2CO+H}
\end{equation}
\begin{equation}
    \ce{CH3O} + \ce{H} \to \ce{CH3OH}.
    \label{eq:CH3O+H}
\end{equation}
\noindent Here, abstraction reactions induced by H atoms also take place, converting \ce{CH3OH} into \ce{CO} through \ce{H2CO} \citep{Hidaka2009,Chuang2016}. However, recent theoretical works that combine a full reaction network for the hydrogenation of CO and microscopic kinetic Monte Carlo simulations of interstellar ices suggest a dominating alternative to reaction (\ref{eq:CH3O+H}), that is the radical-molecule route \citep{Alvarez-Barcia2018,Simons2020}:
\begin{equation}
    \ce{CH3O} + \ce{H2CO} \to \ce{CH3OH} + \ce{HCO},
    \label{eq:CH3O+H2CO}
\end{equation}
\noindent with a contribution of 70--90\% with respect to the remainder of the routes to form methanol. Reaction (\ref{eq:CH3O+H2CO}) can take place upon only one hydrogen-atom addition to \ce{H2CO}, as long as another \ce{H2CO} molecule lies in the vicinity of the formed \ce{CH3O}. Comparatively, reaction (\ref{eq:CH3O+H}) warrants that an additional H atom diffuses through the ice and reacts with \ce{CH3O}. The overall contribution from each route will therefore be dictated by the availability of \ce{H}, \ce{CH3O}, and \ce{H2CO} in the ice. Although the theoretical results represent a significant paradigm change in the final step of the CO hydrogenation formation pathway of methanol in interstellar ices, they have not yet been systematically investigated in the laboratory. Experimental verification is however possible by exploring the kinetic isotope effect of the reactions involving deuterated species (i.e., \ce{D2CO} and \ce{D}), as will be discussed below.

At the typical low temperatures of molecular clouds, quantum-mechanical tunneling governs the activation of chemical reactions. Under these conditions, D-abstraction reactions are significantly hampered in comparison to their H-abstraction counterparts (e.g., \citealt{Nagaoka2005,Hidaka2007,Goumans2011}), on account of the so-called ``kinetic isotope effect''. Reaction (\ref{eq:CH3O+H2CO}) is a radical-molecule route that involves an H-abstraction reaction from \ce{H2CO} with an activation barrier of $\Delta E \sim 2670$ K \citep{Alvarez-Barcia2018}. Thus, it most likely proceeds through quantum-tunneling at such low temperatures. As a consequence, the rate constant of the analog reaction with the deuterated formaldehyde (i.e., D-atom abstraction from $\ce{D2CO}$) is expected to be severely hindered. In the present work, we take advantage of this isotope effect to quantify the respective contributions from the mechanisms (\ref{eq:CH3O+H}) and (\ref{eq:CH3O+H2CO}) in forming methanol under molecular-cloud conditions. We then confirm the dominance of the radical-molecule route through complementary $\ce{H2CO} + \ce{D}$ experiments.


\section{Experimental} \label{sec:exp}

All experiments are performed using the setup SURFRESIDE$^3$, which has been described in detail elsewhere \citep{Ioppolo2013,Qasim2020}. It consists of an ultrahigh vacuum (UHV) chamber with a base pressure of $\sim5\times10^{-10}$ mbar. At its center, a gold-plated copper substrate is mounted on the tip of a closed-cycle He cryostat that allows the temperature of the substrate to vary between 8 and 450 K through resistive heating. The temperature is monitored by two silicon diode sensors with a relative accuracy of 0.5 K. We deposit either \ce{H2CO} or \ce{D2CO} simultaneously with thermalized H and D atoms during the so-called codeposition experiments. The \ce{H2CO} and \ce{D2CO} vapors are produced from paraformaldehyde (purity $95\%$, Sigma-Aldrich) and paraformaldehyde-$d_2$ (purity $98\%$, Sigma-Aldrich) powders heated to $\sim60-100^\circ$ C in glass vacuum tubes. The hydrogen and deuterium atoms are simultaneously generated by both a Hydrogen Atom Beam Source (HABS, \citealt{Tschersich2000}) and a Microwave Atom Source (MWAS, \citealt{Anton2000}), and are thermally cooled to room temperature by colliding with the walls of nose-shaped quartz pipes that are placed along the beam path in both atom sources. A set of experiments comprising of 360 minutes of $\ce{H2CO} + \ce{H}$ or $\ce{D2CO} + \ce{H}$ codepositions is performed at temperatures of 10, 12, 14 and 16 K. Additionally, analogous codeposition experiments of $\ce{H2CO} + \ce{D}$ are also reported. The full complementary set of \ce{H2CO} + \ce{D} and \ce{D2CO} + \ce{D} codepositions are not performed due to the low efficiency of the D-atom addition reactions resulting from the kinetic isotope effect caused by the slower rate of quantum tunneling of D atoms over H atoms \citep{Hidaka2009,Goumans2011}.

During the experiments, the ices are monitored \textit{in situ} by means of Fourier-Transform Reflection-Absorption Infrared Spectroscopy (RAIRS) in the range of 700 to 4000 cm$^{-1}$, with a resolution of 1 cm$^{-1}$. After the codeposition experiments, the sample is heated at a ramping rate of 5 K min$^{-1}$. The desorbed species are monitored by a Quadrupole Mass Spectrometer during Temperature-Programmed Desorption experiments (TPD-QMS). To derive the column densities of the species in the ice, we employ the modified Beer-Lambert law to convert absorbance area to absolute abundance. We use band-strength values of $A(\ce{H2CO})_{\nu=1727}\sim3.6\times10^{-17}$ cm molecule$^{-1}$ and $A(\ce{CH3OH})_{\nu=1030}\sim3.1\times10^{-17}$ cm molecule$^{-1}$, which were calibrated in the same setup by laser-interference measurements \citep{Chuang2018}. For \ce{CHD2OH} ($\nu$=1037 cm$^{-1}$), the band strength is estimated by multiplying $A$(\ce{CH3OH}) by a calibration factor of 0.43 \citep{Nagaoka2007}. Finally, the band strength of \ce{D2CO} is obtained by comparing the IR absorbance areas of the same amount of pure \ce{H2CO} and \ce{D2CO} ices. We experimentally derive $A(\ce{D2CO})_{\nu=1679}\sim2.8\times10^{-17}$ cm molecule$^{-1}$. More details on this procedure are given in Appendix \ref{sec:appendix1}. For a specific flux configuration, the relative uncertainties of the molecule and H-atom fluxes are, respectively, $<$9\% and $<$4\%. The error-bar estimation due to the instrumental uncertainties of the RAIRS and TPD-QMS measurements is derived from the integrated noise signals of blank experiments in the same band width. Table \ref{table:exp_list} in Appendix \ref{sec:appendix2} lists the experiments performed in this study and the relative abundance of \ce{CHD2OH}/\ce{CH3OH} derived from the RAIRS and QMS-TPD analyses.


\section{Results and Discussion} \label{sec:results}

\begin{figure}[htb!]\centering
\includegraphics[scale=0.8]{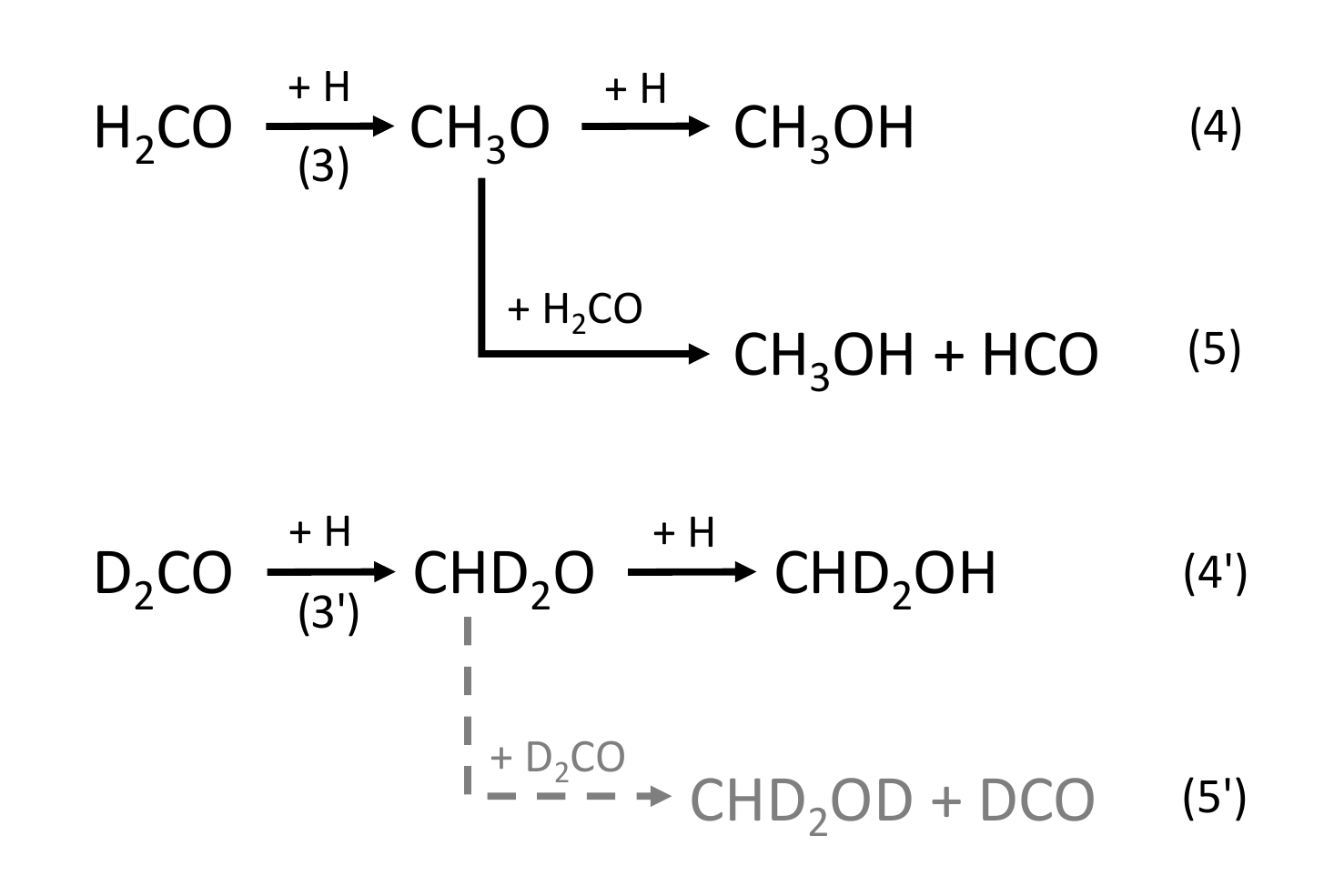}
\caption{\ce{CH3OH} formation scheme involving \ce{H2CO} and \ce{D2CO}. The gray dashed line indicates the smaller efficiency of the deuterated radical-molecule reaction with relation to route (5).}
\label{fig:reaction_scheme}
\end{figure}

In Figure \ref{fig:reaction_scheme}, we present a schematic view of the reaction routes probed in our experiments. To focus on reactions (\ref{eq:CH3O+H}) and (\ref{eq:CH3O+H2CO}), we skip steps (\ref{eq:CO+H}) and (\ref{eq:HCO+H}) and start by directly depositing formaldehyde. This way, comparable starting points between the two scenarios are obtained.

By using different isotope-labeled reactants (\ce{H2CO} and \ce{D2CO}), the importance of reactions (\ref{eq:CH3O+H}) and (4') can be directly assessed. The yields of methanol from \ce{H2CO} through reaction (\ref{eq:CH3O+H}) and from \ce{D2CO} through (4')---see Figure \ref{fig:reaction_scheme}---should be identical, since both radical-radical routes are barrierless. Conversely, the formation of the isotope-labeled \ce{CHD2OD} through reaction (5') is expected to be strongly affected by the isotope effect and to become negligible, whereas (\ref{eq:CH3O+H2CO}) still takes place. Moreover, the hydrogen-addition steps $\ce{H2CO} + \ce{H}$ and $\ce{D2CO} + \ce{H}$ have similar activation barriers of 2318 and 2253 K, respectively, and similar rate constants \citep{Hidaka2009,Goumans2011}. Thus, they yield comparable amounts of \ce{CH3O} and \ce{CHD2O}. The contributions of the two competing routes are hence obtained by comparing the relative abundances between the total \ce{CH3OH} yield of the \ce{H2CO} ice (reactions (\ref{eq:CH3O+H}) and (\ref{eq:CH3O+H2CO})) and that of \ce{CHD2OH} in the \ce{D2CO} ice (reaction (4')).

\begin{figure}[htb!]\centering
\includegraphics[scale=0.5]{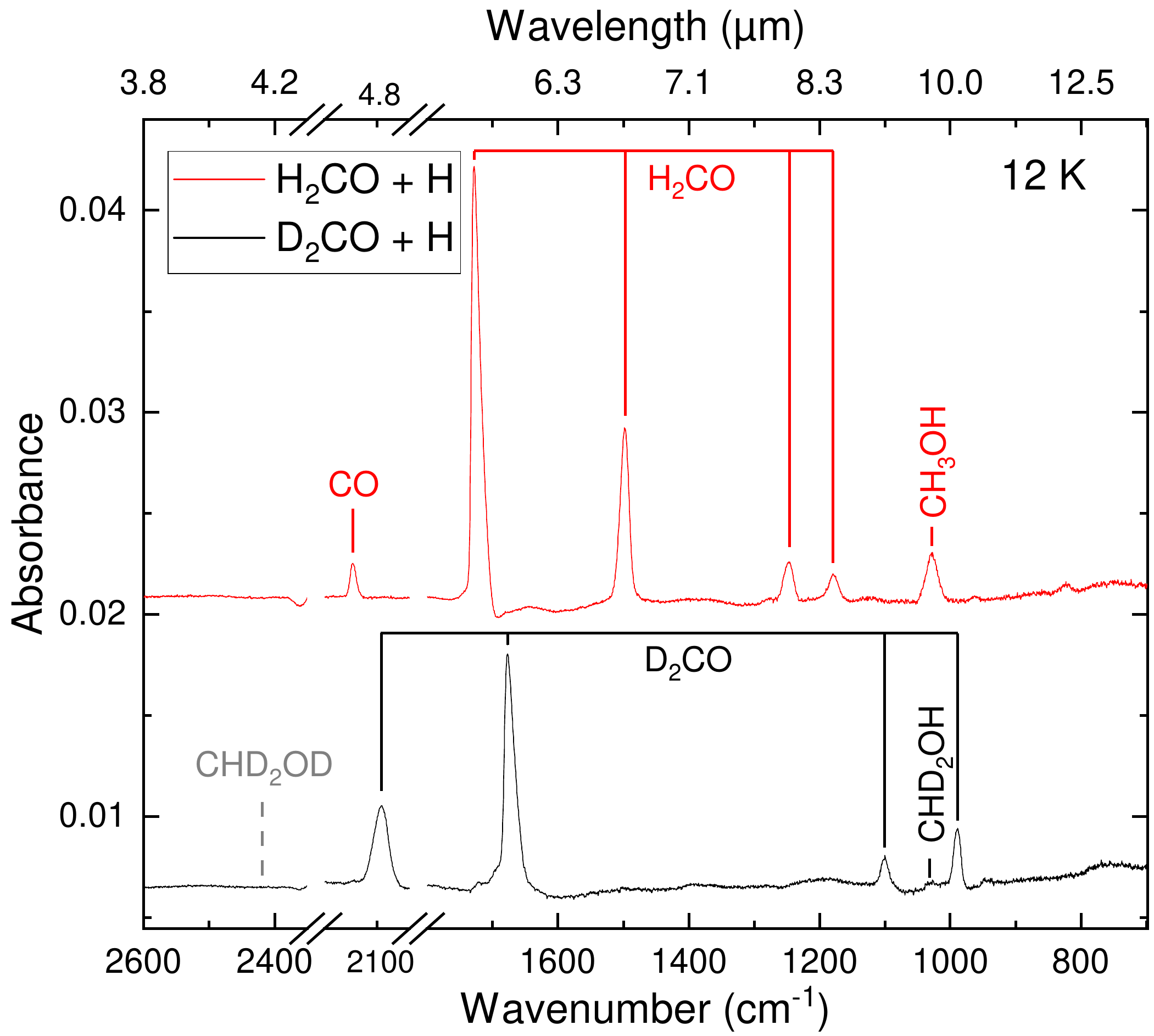}
\caption{Final IR spectra of the $\ce{H2CO} + \ce{H}$ (red) and $\ce{D2CO} + \ce{H}$ (black) codeposition experiments performed at 12 K. The spectra were obtained after 360 minutes of deposition and are artificially offset for clarity. Peak assignments are indicated with solid red (for $\ce{H2CO} + \ce{H}$) and black (for $\ce{D2CO} + \ce{H}$) lines, while the dashed gray line shows the absence of the \ce{CHD2OD} $\nu\approx2420$ cm$^{-1}$ peak.}
\label{fig:IR_spectra_12K}
\end{figure}

The final infrared spectra of both $\ce{H2CO} + \ce{H}$ and $\ce{D2CO} + \ce{H}$ codepositions after 360 minutes at 12 K are shown in Figure \ref{fig:IR_spectra_12K}. IR features of \ce{H2CO} are observed at 1167 (\ce{CH2} wag., $\nu=6$), 1249 (\ce{CH2} rock., $\nu=5$), 1500 (\ce{CH2} scis., $\nu=3$) and 1727 cm$^{-1}$ (\ce{CO} str., $\nu=2$). In the analog deuterated experiments, \ce{D2CO} peaks are detected at 988 (\ce{CD2} wag.), 1101 (\ce{CD2} scis.) and 1679 cm$^{-1}$ (\ce{CO} str.). In both cases the hydrogenated products \ce{CH3OH} and \ce{CHD2OH} are detected by their \ce{CO} stretching ($\nu=8$) features at 1030 and 1037 cm$^{-1}$, respectively \citep{Shimanouchi1972,Hidaka2009}, which indicates that the H-addition routes in Figure \ref{fig:reaction_scheme} proceed at a considerable rate.

Even in the experimental configuration with the highest product yield, no \ce{CHD2OD} peaks (OD str. at $\nu_1\sim2420$ cm$^{-1}$, \citealt{Nagaoka2007}) could be observed in the $\ce{D2CO} + \ce{H}$ spectrum. The detection of \ce{CHD2OH} and non-detection of \ce{CHD2OD} are in line with the expected low efficiency of the radical-molecule route for deuterated species and are consistent with the results by \cite{Hidaka2009} on the exposure of H atoms to \ce{D2CO} ice. That, in addition to the fact that the \ce{CHD2OH} peak is much smaller than the \ce{CH3OH} counterpart in the $\ce{H2CO} + \ce{H}$ experiment at equivalent conditions, strongly suggests that reaction (\ref{eq:CH3O+H}) is not the dominant route to form methanol under the investigated experimental conditions. Instead, reaction (\ref{eq:CH3O+H2CO}) must play a more prominent role in the formation of methanol. This observation remains valid after converting the integrated peak areas from the RAIR spectra to the species' abundances, as shown in Table \ref{table:exp_list}. From the RAIRS analyses, we do not detect any bands related to the radicals \ce{HCO} or \ce{CH2OH}---nor their deuterated isotopes---reinforcing the findings from previous studies that their high reactivity results in low abundances \citep{Watanabe2002,Fuchs2009}. The radical \ce{CH3O} is also not observed in the \ce{H2CO} + \ce{H} experiments. It should be noted, though, that the most intense IR features of its isotopologue \ce{CHD2O} overlap with those of the other reactants and products in the spectra (see, e.g., \citealt{Haupa2017}), and therefore we cannot rule out its presence in the \ce{D2CO} + \ce{H} experiments. Moreover, we observe a clear \ce{CO} peak ($\nu=2138$ cm$^{-1}$) in the $\ce{H2CO} + \ce{H}$ experiment, as well as \ce{HDCO} ($\nu=1694$ cm$^{-1}$) and \ce{H2CO} ($\nu=1727$ cm$^{-1}$) peaks in the $\ce{D2CO} + \ce{H}$ counterpart, in agreement with the results by, e.g., \cite{Hidaka2009} and \cite{Chuang2016} that H-atom induced abstraction reactions take place for the studied conditions (see, e.g., Figure 5 in \citealt{Hidaka2009}).

\begin{figure}[htb!]\centering
\includegraphics[scale=0.5]{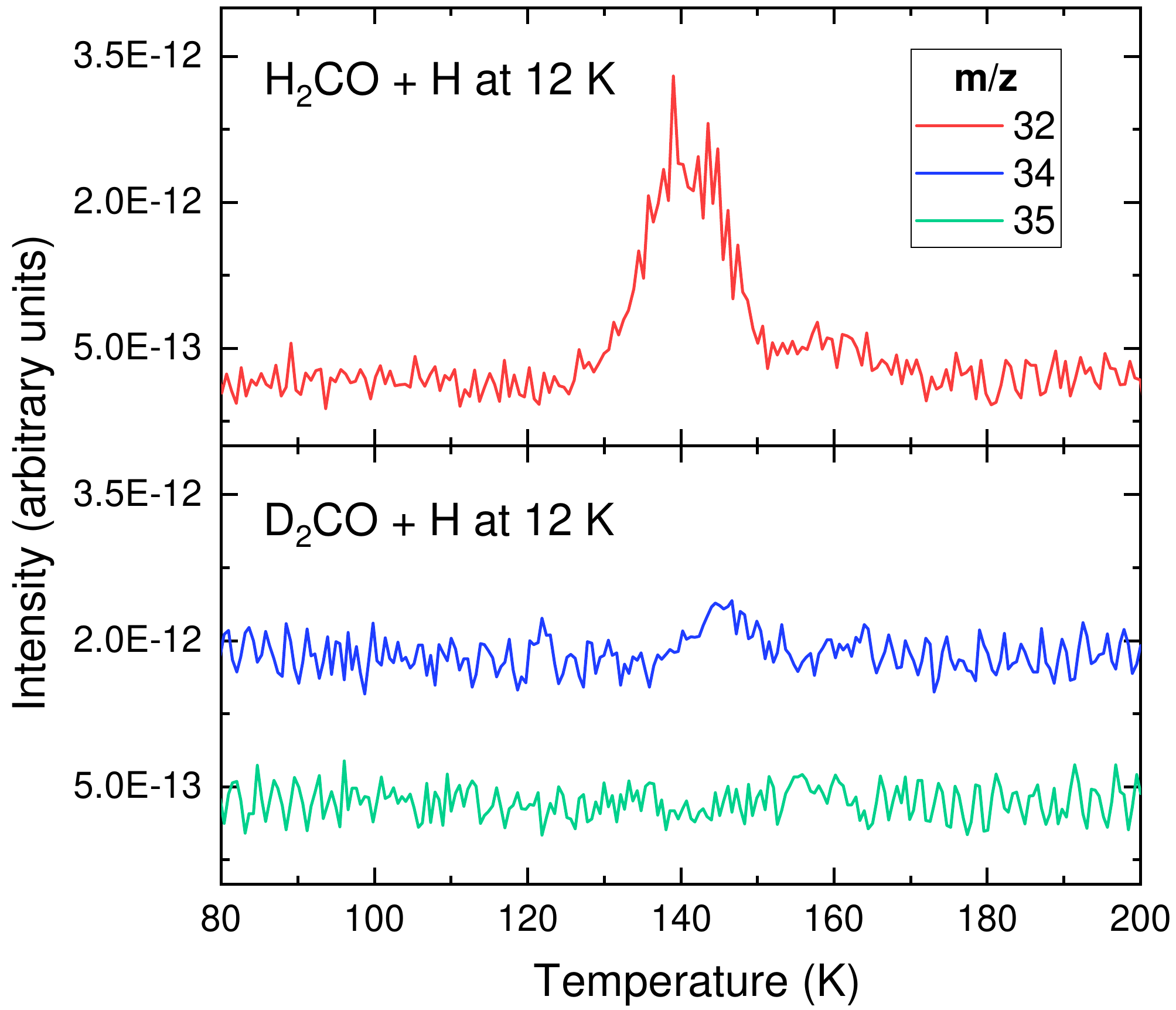}
\caption{Spectra of the TPD-QMS experiments obtained after codeposition of $\ce{H2CO} + \ce{H}$ (upper) and $\ce{D2CO} + \ce{H}$ (lower) at 12 K for 360 min. The \textit{m/z} signals of 32 (red), 34 (blue) and 35 (green) correspond to, respectively, \ce{CH3OH}, \ce{CHD2OH} and \ce{CHD2OD}. No signal is detected for m/z = 35 at the desorption temperature of methanol ($\sim$140 K). }
\label{fig:tpd_12K}
\end{figure}

The TPD-QMS results of the experiments at 12 K are presented in Figure \ref{fig:tpd_12K} as a typical example of the obtained spectra. Since trying to account for the mass-fragmentation patterns induced by electron impact may complicate the interpretation of the data, only the ion signals of the products' molecular masses are selected to quantify their formation yield. For example, the \ce{CH3OH} formation is quantified by the mass signal m/z = 32, while the deuterated species \ce{CHD2OH} and \ce{CHD2OD} are depicted by m/z = 34 and 35, respectively. All methanol (-$d_0$--$d_3$) mass signals peak at $\sim$140 K. The small m/z = 32 signal that peaks at 160 K could be due to either \ce{CH3OH} trapped in background water or fragments originating from complex organic molecules \citep{Fedoseev2015,Chuang2016,He2021}. The flux of background water deposited throughout the experiment is estimated to be 5.2$\times$10$^{10}$ molecules cm$^{-2}$ s$^{-1}$, being a minor component of the ice that does not affect the final results. The TPD spectra show that a much smaller yield of \ce{CHD2OH} is observed in comparison to \ce{CH3OH}, and no \ce{CHD2OD} is detected above the instrumental detection limit, thereby confirming the RAIRS analysis.

\begin{figure}[htb!]\centering
\includegraphics[scale=0.5]{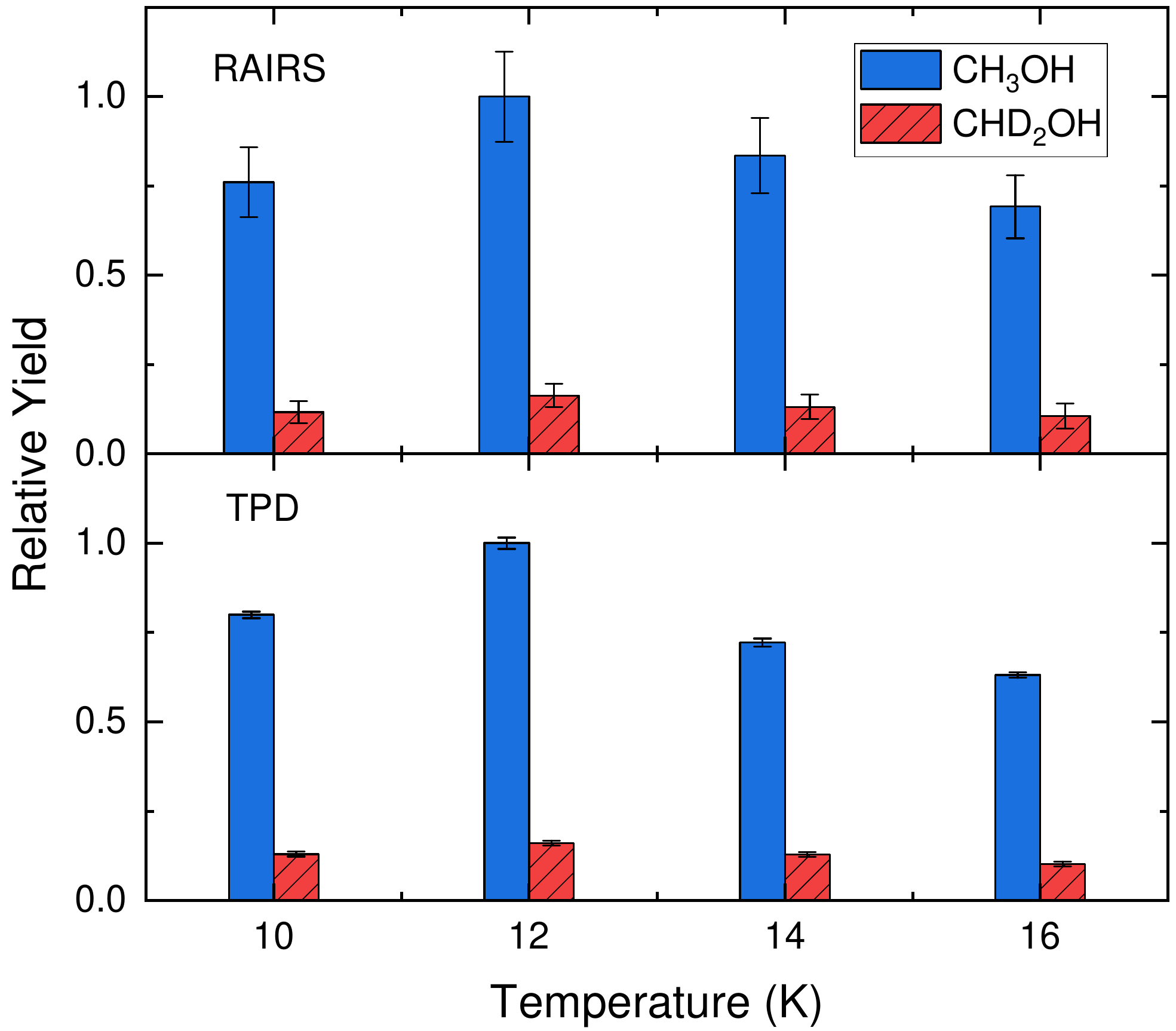}
\caption{\textit{Upper panel:} Yields of \ce{CH3OH} and \ce{CHD2OH} relative to the column density of \ce{CH3OH} after 360 minutes of $\ce{H2CO} + \ce{H}$ codeposition at 10, 12, 14, and 16 K, as derived from the RAIRS data. \textit{Lower panel:} Same as the upper panel, but for the TPD-QMS data. The reported values are normalized to the most abundant yield at 12 K.}
\label{fig:RAIRS_TPD_hist}
\end{figure}

Figure \ref{fig:RAIRS_TPD_hist} presents the relative yields of \ce{CH3OH} and \ce{CHD2OH} derived from the RAIRS and QMS-TPD analyses at 10, 12, 14 and 16 K. The methanol yield of the hydrogenation reactions varies with the surface temperature, showing the effective result of the competition between the increase in the diffusion of H atoms and the decrease in their residence time on the ice as a function of temperature from 10 to 16 K. The highest abundances of methanol in both experiments $\ce{H2CO} + \ce{H}$ and $\ce{D2CO} + \ce{H}$ are found at 12 K, and drop with increasing temperatures. A set of experiments performed at 20 K results in \ce{CHD2OH} yields that are not measurable, and thus is not reported here. Therefore, the abundances are normalized to that of \ce{CH3OH} at 12 K. Similar observations have been reported in a previous study on the hydrogenation of \ce{H2CO} at temperatures up to 25 K \citep{Chuang2016}.

For the entire set of experiments, the RAIRS data show a considerably lower abundance of \ce{CHD2OH} from the $\ce{D2CO} + \ce{H}$ experiments compared to that of \ce{CH3OH} from $\ce{H2CO} + \ce{H}$. In fact, the abundance ratios of \ce{CHD2OH}/\ce{CH3OH} between analogous experiments yield a constant value of $\sim$ 0.16 for the temperature range between 10 and 16 K (see Table \ref{table:exp_list}). This is confirmed by the QMS-TPD data, which agree well within their uncertainties with the results from the RAIRS analysis (\ce{CHD2OH}/\ce{CH3OH} $\sim$ 0.17). The contributions of the reactions probed here are therefore independent of these temperatures (10--16 K), which are of relevance to molecular clouds. This observation is in accordance with the predictions of the models by \citet{Simons2020}. The similarity between both the RAIRS and TPD ratios also excludes the possibility of the former being influenced by the heating of the substrate.

Under the assumption that reactions (\ref{eq:H2CO+H}) and (3') result in similar amounts of \ce{CH3O} and \ce{CHD2O} respectively, and since both reactions (\ref{eq:CH3O+H}) and (4') are barrierless, the methanol yields from the abstraction route should be similar in both $\ce{H2CO} + \ce{H}$ and $\ce{D2CO} + \ce{H}$ experiments;
\begin{equation}
    (4) \sim (4'),
\end{equation}
\noindent where the parentheses denote the methanol yield of the reactions. The discrepancy between the yields of \ce{CH3OH} and \ce{CHD2OH} under our experimental conditions can therefore be used to estimate the contribution $C4$ of reaction (\ref{eq:CH3O+H}) with respect to the total yield of methanol:
\begin{equation}
    \begin{split}
    C4=\frac{(4)}{(4) + (5)}\sim\frac{(4')}{(4) + (5)}\\
    \sim\frac{(\ce{CHD2OH})}{(\ce{CH3OH})}\sim[0.16-0.17].
    \end{split}
\end{equation}
The estimation described here assumes that the formation of \ce{CHD2OH} and \ce{CH3OH} takes place predominantly through the reactions presented in Figure \ref{fig:reaction_scheme}. Other routes to form methanol, such as $\ce{CH2OH} + \ce{H}$, have very minor contributions and are hence disregarded based on the inefficient formation of \ce{CH2OH} from \ce{H2CO} hydrogenation (\citealt{Song:2017} and, e.g., Table 6 in \citealt{Simons2020}). Accordingly, the contribution $C5$ of reaction (\ref{eq:CH3O+H2CO}) corresponds to:
\begin{equation}
    C5 = 1 - C4 \sim [0.83-0.84].
\end{equation}
The experimentally derived value of $C5$ agrees well with the contribution of $\sim$90\% reported by \cite{Simons2020} from kinetic Monte Carlo simulations of the hydrogenation of \ce{H2CO} ices.

The \ce{H}/\ce{H2CO} flux ratios employed here were chosen based on the conditions of the models by \cite{Simons2020} (\ce{H}:\ce{H2CO} = 20:1). A higher H flux could potentially favor the atom-addition route and thus lead to a lower product ratio. To verify whether the contributions $C4$ and $C5$ depend on the hydrogen-to-molecule ratio, a set of experiments using \ce{H}/\ce{H2CO}(\ce{D2CO})$=$30 was performed at 10 K (see Table \ref{table:exp_list}). The resulting \ce{CHD2OH}/\ce{CH3OH} ratios derived from the RAIRS and TPD-QMS data are 0.18$\pm$0.03 for $C4$ and 0.82$\pm$0.03 for $C5$. These values are negligibly different considering the detection error. Thus, a higher hydrogen flux affects only slightly the contributions derived above, and reaction (\ref{eq:CH3O+H2CO}) still governs the formation of \ce{CH3OH}. This implies that both flux ratios employed here represent an overabundance of H atoms, and therefore the product yield is limited by either \ce{H2CO} or \ce{D2CO}.

In our experiments, the dominance of the radical-molecule route is likely due to the higher availability of \ce{H2CO} in the environment, as opposed to H atoms, to react with \ce{CH3O}. Both \ce{CH3O} and \ce{H} are very reactive species; when a \ce{CH3O} radical is formed in the vicinity of a \ce{H2CO} molecule, it readily reacts with \ce{H2CO} to form \ce{CH3OH} + \ce{HCO}. On the other hand, the accreted H atoms on the surfaces are required to diffuse in order to react with the \ce{CH3O} radicals. Therefore, they will mostly react with the broadly available \ce{H2CO}, and less likely with \ce{CH3O}, which is a minor ice component. Moreover, the diffusion of hydrogen atoms also competes with its reaction with \ce{CH3O}, whereas both \ce{CH3O} and \ce{H2CO} do not diffuse and therefore have more time available to react. As a result, the higher likelihood of the two reactants in reaction (\ref{eq:CH3O+H2CO}) to meet overpowers reaction (\ref{eq:CH3O+H}), despite the latter being barrierless. 

Although the ice composition explored here is not fully representative of interstellar ices, the foregoing reaction routes are nonetheless relevant to their chemistry. In more realistic ices, there is a lower availability of formaldehyde, and the formed \ce{CH3O} radicals can also react with other species. Nevertheless, the aforementioned reasoning still applies, since the H atoms will mostly react with \ce{CO} and \ce{HCO}, additionally to \ce{H2CO}. Indeed, route (\ref{eq:CH3O+H2CO}) was observed to dominate the formation of methanol even in simulations starting from the hydrogenation of \ce{CO} ice \citep{Simons2020}. Moreover, the radical-molecule route forms \ce{HCO} as a byproduct, which will then be hydrogenated again to \ce{H2CO}, replenishing and thus favoring the chemical network. Additional formaldehyde formed in the gas phase in the ISM can also subsequently accrete onto ice grains and take part in the radical-molecule route. 

It is noteworthy that the hydrogenation of pure \ce{D2CO} ice is not perfectly equivalent to that of pure \ce{H2CO} ice. The rate constant of reaction (3') is in fact slightly higher than that of reaction (\ref{eq:H2CO+H}) \citep{Hidaka2009, Goumans2011}. Moreover, reaction (5') is inhibited by the isotope effect. Thus, there should be an increase in the availability of \ce{CHD2O} in the ice during the deuterated experiments, which in turn favors the formation of \ce{CHD2OH} through route (4'). The surplus of \ce{CHD2O} is either consumed through H-induced abstraction reactions or preserved in the ice. Furthermore, the isotope effect also hampers the D abstraction reaction $\ce{D2CO} + \ce{H} \to \ce{DCO} + \ce{HD}$, which is why the formation of \ce{CO} is much less efficient in these scenarios (see Figure \ref{fig:IR_spectra_12K}). This results in a higher availability of H atoms in the surface of the ice in the deuterated experiments, compared to the non-deuterated counterparts where the abstraction reaction $\ce{H2CO} + \ce{H} \to \ce{HCO} + \ce{H2}$ is faster. As a consequence, the formation of \ce{CHD2OH} is also favored by the surplus of hydrogen atoms. Therefore, the contribution from reaction (4') to forming methanol under the present experimental conditions should be regarded as an upper limit to that of reaction (\ref{eq:CH3O+H}), which puts a lower limit to reaction (\ref{eq:CH3O+H2CO}) as being at least roughly four times more efficient than reaction (\ref{eq:CH3O+H}).

Aiming to directly confirm the occurrence of the radical-molecule route forming methanol, an experiment of $\ce{H2CO}+\ce{D}$ and a control experiment of D-atom beam exposure to the bare substrate are performed at 10 K for six hours. The codeposition experiment is followed by a TPD with a ramping rate of 5 K/min. These experiments are designed to probe the following competing reactions:
\begin{equation}
    \ce{H2CO} \xrightarrow{+\ce{D}} \ce{CH2DO} \xrightarrow{+\ce{D}} \ce{CH2DOD},
\label{eq:D_add}    
\end{equation}
\begin{equation}
    \ce{H2CO} \xrightarrow{+\ce{D}} \ce{CH2DO} \xrightarrow{+\ce{H2CO}} \ce{CH2DOH} + \ce{HCO}.
\label{eq:D_subs}    
\end{equation}

\begin{figure*}[htb!]\centering
\includegraphics[scale=0.77]{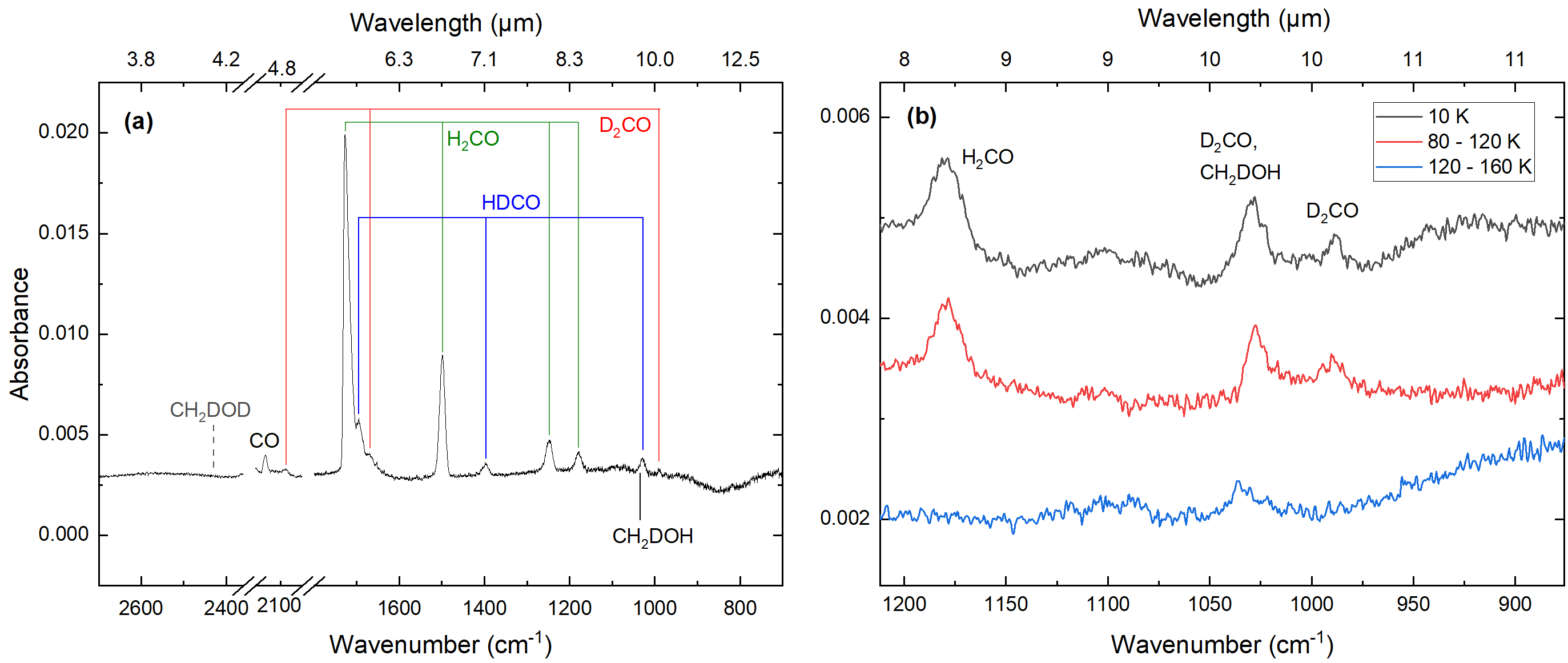}
\caption{(a) Difference spectrum of the final spectra acquired after 360 minutes of \ce{H2CO} + \ce{D} codeposition minus that of the analogous \ce{D}-atom beam deposition for the same time, both performed at 10 K. The dashed line indicates the absence of the \ce{CH2DOD} feature at $\nu\sim2430$ cm$^{-1}$. (b) Infrared spectra obtained during the ramp-up after the \ce{H2CO} + \ce{D} 360-minutes codeposition. The gray, red and blue spectra were measured at 10 K, $80-120$ K, and $120-160$ K, respectively.}
\label{fig:H2CO+D}
\end{figure*}

The difference of the spectra acquired after the \ce{H2CO} + \ce{D} experiment and the \ce{D} deposition is shown in panel (a) of Figure \ref{fig:H2CO+D}. In agreement with \cite{Hidaka2009}, we observe spectral features of \ce{HDCO} and \ce{D2CO} as a result of D-substitution (abstraction + addition) reactions on \ce{H2CO}. Moreover, a signal at $\nu$ = 1034 cm$^{-1}$ is in line with the CO stretching band of \ce{CH2DOH}, which would indicate that the radical-molecule route (i.e., reaction (\ref{eq:D_subs})) proceeds. However, this feature is (at least partially) blended with the CH$_2$ rocking mode of \ce{HDCO} ($\nu$ = 1029 cm$^{-1}$), to an extent where deconvolution is not possible. Nonetheless, given the lower desorption temperature of \ce{HDCO} ($\sim$100 K) with respect to that of \ce{CH2DOH} ($\sim$140 K), it is possible to isolate the contribution from \ce{CH2DOH} by analyzing the infrared spectra obtained during the TPD experiment. The spectra acquired at 10 K and in the temperature intervals of $80-120$ K and $120-160$ K are shown in panel (b) of Figure \ref{fig:H2CO+D}.

In the IR spectrum acquired in the temperature range of $80-120$ K, the peak at $\nu \sim$1029 cm$^{-1}$ contains the contributions from both \ce{HDCO} and \ce{CH2DOH}. However, at the interval of $120-160$ K, the peak slightly shifts to $\nu \sim$1034 cm$^{-1}$, and only accounts for the \ce{CH2DOH} left on the substrate. Based on the obtained area of this peak and its reported band strength \citep{Nagaoka2007}, we derive a \ce{CH2DOH} column density of $\sim(3.0\pm0.3)\times10^{14}$ molecules cm$^{-2}$. This yield is $\sim4$ times smaller than that of \ce{CH3OH} from the \ce{H2CO} + \ce{H} experiment at 10 K (\ce{H}/\ce{H2CO} = 10), which is consistent with a lower rate for the D-addition reaction to \ce{H2CO} over the H-addition counterpart.

During TPD, the mass signal m/z = 33 is detected at $\sim$150 K. However, whereas the spectroscopic signals are quite unambiguous, this does not apply to the mass spectrometric data. Since the mass signals of \ce{CH2DOH} (m/z = 33, 32, 31...) overlap with those of \ce{CH2DOD} (m/z = 34, 33, 32, …), and since their fragmentation patterns are unknown, the interpretation of the mass spectrometry data is impeded.

The product of the D-addition route, \ce{CH2DOD} (reaction (\ref{eq:D_add})), was not observed in its OD stretching mode ($\nu\sim2430$ cm$^{-1}$) within the detection limit of the instrument, as shown by the dashed line in panel (a) of Figure \ref{fig:H2CO+D}. Given that the band strength of this feature is expected to be around 1.5 times that of the CO stretching mode of \ce{CH2DOH} \citep{Nagaoka2007}, the abundance of \ce{CH2DOD} must be appreciably smaller than that of \ce{CH2DOH}. This result provides additional compelling evidence that the radical-molecule route dominates the formation of methanol under the present conditions.

\section{Conclusions} \label{sec:conclusions}

In the present work, we experimentally examine the contribution from the conventional route to the formation of methanol (reaction (\ref{eq:CH3O+H})) with reaction (\ref{eq:CH3O+H2CO}), and provide the respective impact of the two proposed last steps of relevance to the chemistry of molecular clouds. This was done by focusing on the specific reaction step in question, as more realistic experiments with \ce{CO} ices would be much more difficult to interpret. Our main conclusions are the following:

\begin{itemize}
\item We derive a contribution of $\gtrsim$83\% for the radical-molecule route (\ref{eq:CH3O+H2CO}), which is independent of the \ce{H}/\ce{H2CO} ratios (i.e., 10 and 30) and the temperatures (i.e., $10-16$ K) experimentally investigated in this work. This temperature range is typical of dense clouds, where \ce{H} atoms are expected to be overabundant in comparison with \ce{H2CO}. Our experiments confirm the results in \cite{Simons2020}, who performed simulations for conditions as used in the laboratory work presented here. Their model, in turn, extends to the interstellar medium, and concludes that this route dominates the formation of interstellar methanol. Our results corroborate with this finding. Comparatively, reaction (\ref{eq:CH3O+H}) should account for $\lesssim$17\% of the methanol formation under the present conditions.

\item Additional codeposition experiments of \ce{H2CO} and \ce{D} yield measurable amounts of \ce{CH2DOH}, whereas \ce{CH2DOD} is not detectable within the IR experimental limits. This further confirms the dominance of the radical-molecule route to form methanol under our experimental conditions.

\item The conclusion drawn here does not affect the consensus that methanol is mainly formed through the hydrogenation of \ce{CO} in the solid phase, albeit by means of a different mechanism than originally thought. Nonetheless, this finding represents an important change in the perspective of methanol formation in space, and could affect astrochemical models that involve this key species.

\item The expected deuterium fractionation of methanol will likely be affected, with a possibly smaller abundance of deuterated species due to the kinetic isotope effect.

\item Furthermore, the \ce{HCO} radical is an important astrochemical intermediate since it can react to further form \ce{H2CO} or recombine with other radicals (e.g., \ce{CH3O} and \ce{CH2OH}) to produce complex organic molecules (COMs). The alternative route probed here forms \ce{HCO} radicals as a byproduct, which enriches an icy surface with this species and therefore affects the COM distribution in interstellar ices, likely enhancing the predicted formation rate of, e.g., glycolaldehyde and ethylene glycol in previous astrochemical models that did not consider reaction (\ref{eq:CH3O+H2CO}) as a dominant channel. Models of the radical-radical reactions driven by the warming of the ice at later stages of star-formation will also be affected by the additional HCO that eventually remains preserved in the ice and that becomes mobile and reacts.

\item The inclusion of deuterium atoms to microscopic kinetic Monte Carlo simulations would be a logical next step, as it could provide additional insight on the chemical network involving methanol. This, however, is beyond the scope of the present work.
\end{itemize}

\begin{acknowledgments}
\section*{Acknowledgments}
The authors would like to thank Herma Cuppen for many guiding scientific discussions on the mechanism in question.
This work has been supported by the Danish National Research Foundation through the Center of Excellence “InterCat” (Grant agreement no.: DNRF150). It has also been funded by the Dutch Astrochemistry Network II (DANII) and NOVA (the Netherlands Research School for Astronomy).  
T.L. is grateful for support from NWO via a VENI fellowship (722.017.008). G.F. acknowledges financial support from the Russian Ministry of Science and Higher Education via the State Assignment Contract FEUZ-2020-0038. S.I. acknowledges the Royal Society for financial support.

\end{acknowledgments}

\appendix

\section{\ce{D2CO} band strength estimation} \label{sec:appendix1}

\begin{figure}[htb!]\centering
\includegraphics[scale=0.5]{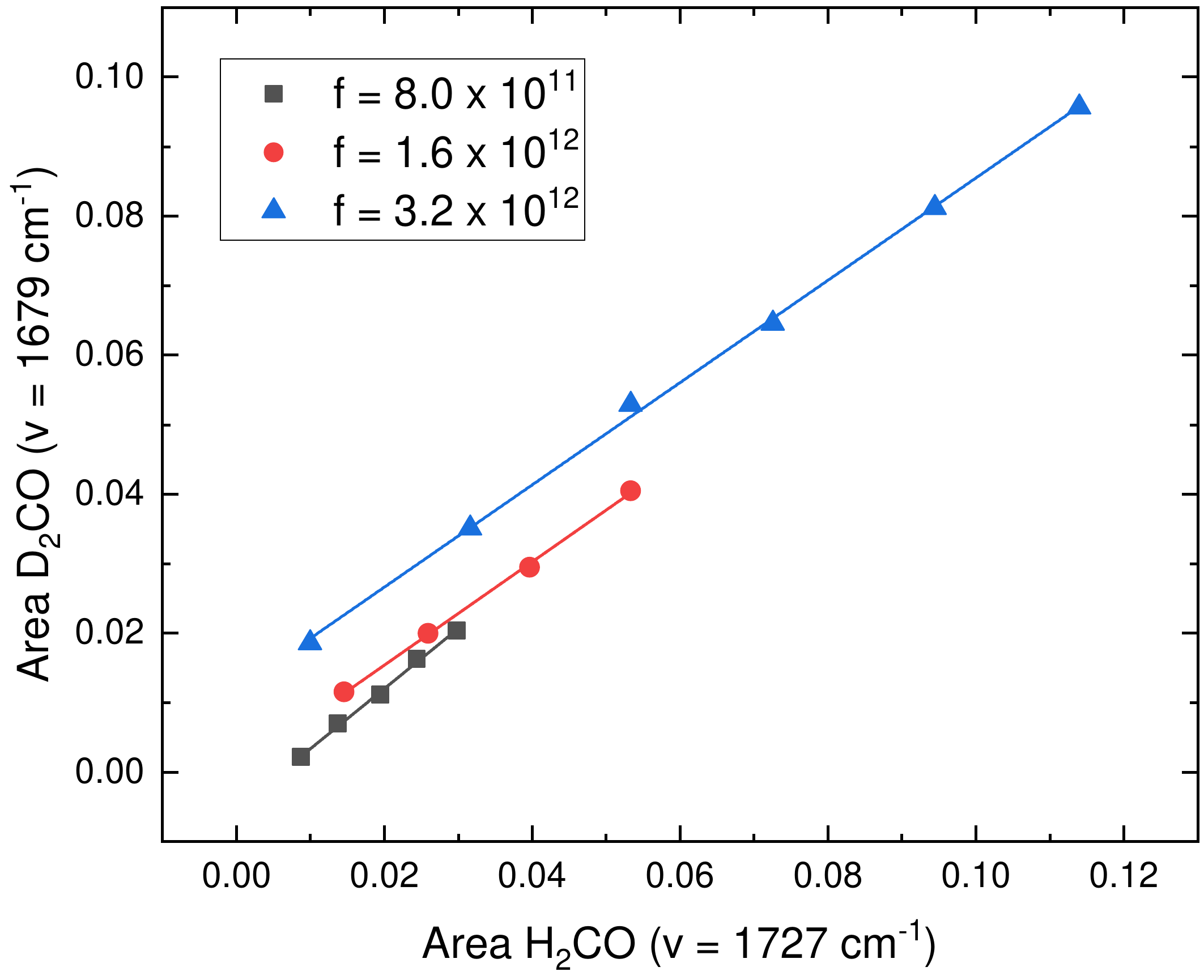}
\caption{Plot of the integrated area of the \ce{D2CO} ($\nu_2=1679$ cm$^{-1}$) infrared absorption band versus that of \ce{H2CO} ($\nu_2=1727$ cm$^{-1}$) for three different molecule fluxes, f: $8\times10^{11}$, $1.6\times10^{12}$, and $3.2\times10^{12}$ cm$^{-2}$ s$^{-1}$.}
\label{fig:band_str_cal}
\end{figure}

The column densities ($N_X$) of the species in the ice are calculated by means of the modified Beer-Lambert equation:
\begin{equation}
    N_X = \ln{10}\frac{\int Abs(\nu)d\nu}{A_X},
    \label{eq:mod_lamb_beer}
\end{equation}
\noindent in which $\int Abs(\nu) d\nu$ is the integrated absorbance of a given infrared band and $A_X$ is its corresponding band strength. To estimate the band strength of \ce{D2CO}, the same deposition conditions (i.e., exposure time and dose) are applied to both pure \ce{H2CO} and \ce{D2CO}, ensuring identical abundances of both types of ices. This is achieved by using the same leak valve while keeping the same flow-rate setting. Previously, \cite{Nagaoka2007} used this method to determine the relative integrated band strengths of \ce{CH3OH} and isotopes. We plot the integrated areas of the \ce{D2CO} $\nu_2=1679$ cm$^{-1}$ band versus those of \ce{H2CO} $\nu_2=1727$ cm$^{-1}$ for different accumulated abundances performed at three different flux configurations, as shown in Figure \ref{fig:band_str_cal}.

After performing a linear fit to the points in the plot, we derive an average slope of $a=0.78\pm0.03$. This slope corresponds to the conversion factor between $A$(\ce{H2CO}) and  $A$(\ce{D2CO}):
\begin{equation}
    A(\ce{D2CO})_{\nu2}=a\times A(\ce{H2CO})_{\nu2}
    \label{eq:prop_A}
\end{equation}
Given that $A$(\ce{H2CO}) for the $\nu=1727$ cm$^{-1}$ peak has been previously measured to be $\sim3.6\times10^{-17}$ cm molecule$^{-1}$ \citep{Chuang2018}, we estimate the \ce{D2CO} $\nu=1679$ cm$^{-1}$ band strength to be $A(\ce{D2CO})\sim2.8\times10^{-17}$ cm molecule$^{-1}$.

\section{Experiment list} \label{sec:appendix2}

\begin{table}[htb!]
\centering
\caption{Overview of the experiments performed in this work, as well as the resulting \ce{CHD2OH}/\ce{CH3OH} abundance ratios between each set of experiments derived from the RAIRS and TPD-QMS analyses. The former was calculated from the CO stretching bands ($\nu$=8) of \ce{CHD2OH} and \ce{CH3OH}, and the latter from the m/z = 32 and m/z = 34 mass signals.}
\label{table:exp_list}      
\begin{tabular}{ccccccc}  
\toprule\midrule
Experiment              &   T$_{\text{sample}}$ &   Molecule Flux                       &   H (D) flux                              &   Time    &   \ce{CHD2OH}/\ce{CH3OH}              &   \ce{CHD2OH}/\ce{CH3OH}\\
                        &   (K)                 &   (cm$^{-2}$ s$^{-1}$)                &   (cm$^{-2}$ s$^{-1}$)                &   (min)   &   (RAIRS, $\nu=8/\nu=8)$                             & (TPD)\\                                               
\midrule
$\ce{D2CO} + \ce{H}$    &   10                  &   $6\times10^{11}$                    &   $6\times10^{12}$                    &   360     &   \multirow{2}{*}{$0.15\pm0.04$}      &   \multirow{2}{*}{$0.16\pm0.01$}\\
$\ce{H2CO} + \ce{H}$    &   10                  &   $6\times10^{11}$                    &   $6\times10^{12}$                    &   360     &                                       &\\
                        &                       &                                       &                                       &           &                                       &\\
$\ce{D2CO} + \ce{H}$    &   12                  &   $6\times10^{11}$                    &   $6\times10^{12}$                    &   360     &   \multirow{2}{*}{$0.16\pm0.04$}      &   \multirow{2}{*}{$0.16\pm0.01$}\\
$\ce{H2CO} + \ce{H}$    &   12                  &   $6\times10^{11}$                    &   $6\times10^{12}$                    &   360     &                                       &\\
                        &                       &                                       &                                       &           &                                       &\\
$\ce{D2CO} + \ce{H}$    &   14                  &   $6\times10^{11}$                    &   $6\times10^{12}$                    &   360     &   \multirow{2}{*}{$0.16\pm0.04$}      &   \multirow{2}{*}{$0.18\pm0.01$}\\
$\ce{H2CO} + \ce{H}$    &   14                  &   $6\times10^{11}$                    &   $6\times10^{12}$                    &   360     &                                       &\\
                        &                       &                                       &                                       &           &                                       &\\
$\ce{D2CO} + \ce{H}$    &   16                  &   $6\times10^{11}$                    &   $6\times10^{12}$                    &   360     &   \multirow{2}{*}{$0.15\pm0.04$}      &   \multirow{2}{*}{$0.16\pm0.01$}\\
$\ce{H2CO} + \ce{H}$    &   16                  &   $6\times10^{11}$                    &   $6\times10^{12}$                    &   360     &                                       &\\
                        &                       &                                       &                                       &           &                                       &\\
$\ce{D2CO} + \ce{H}$    &   10                  &   $6\times10^{11}$                    &   $1.8\times10^{13}$                  &   360     &   \multirow{2}{*}{$0.18\pm0.03$}      &   \multirow{2}{*}{$0.18\pm0.01$}\\
$\ce{H2CO} + \ce{H}$    &   10                  &   $6\times10^{11}$                    &   $1.8\times10^{13}$                  &   360     &                                       &\\
                        &                       &                                       &                                       &           &                                       &\\
$\ce{H2CO} + \ce{D}$    &   10                  &   $4\times10^{11}$                    &   $6\times10^{12}$                    &   360     &                                       &\\
\midrule\bottomrule
\end{tabular}
\end{table}


\bibliography{mybibfile}{}
\bibliographystyle{aasjournal}



\end{document}